\documentclass{appolb}
\usepackage{graphicx}
\usepackage{wrapfig}

\begin{document}
\title{The meson-exchange induced light--by--light contribution to $(g-2)_\mu$ within the nonlocal chiral quark model  %
\thanks{Presented by A.E.D. at Light Cone Conference, 8-13 July 2012, Cracow, Poland.}%
}
\author{A.~E.~Dorokhov
\address{Bogoliubov Laboratory of Theoretical Physics, JINR, 141980 Dubna, Russia
          \and
          N.N.Bogoliubov Institute of Theoretical Problems of Microworld, M.V.Lomonosov Moscow State University, Moscow 119991, Russia}\\[0.3cm]
{A.~E.~Radzhabov, A.~S.~Zhevlakov}
\address{Institute for System Dynamics and Control Theory SB RAS, 664033 Irkutsk, Russia}
}
\maketitle
\begin{abstract}
The current status of the muon anomalous magnetic moment (AMM) problem is briefly
presented. The corrections to the muon AMM coming from the effects of hadronic
light-by-light (LbL) scattering due to light quark-antiquark exchange in pseudoscalar and scalar channels are estimated within the nonlocal chiral quark model (N$\chi$QM). Within this approach the full kinematic dependence on the photon and meson virtualities are taken into account. As a result, the meson exchange contributions to the muon AMM calculated within N$\chi$QM are, in general, smaller than the contributions obtained within other models. The contribution from the scalar channel is positive and small, but stabilizes the combined with pseudoscalar channel result with respect to variation of the model parameters.

\end{abstract}

\section{Muon AMM: theory vs experiment}

The anomalous magnetic moment of the muon is known to an unprecedented
accuracy. The latest result from the measurements of the
Muon $(g-2)$ collaboration at Brookhaven is \cite{Bennett:2006fi}
\begin{equation}
a_{\mu}^{\mathrm{BNL}}=11\ 659\ 208.0\left( 6.3\right)
\cdot10^{-10},  \label{AMMg-2}
\end{equation}
which is a 0.54 ppm uncertainty over combined positive and negative muon measurements.
Using $e^{+}e^{-}$ annihilation and inclusive hadronic $\tau$ decay data,
the standard model predicts \cite{Davier:2010nc}
\begin{equation}
a_{\mu}^{SM}=\left\{
\begin{array}[l]{l}
11\ 659\ 180.2\left( 4.9\right) \cdot10^{-10},\qquad [e^{+}e^{-}], \\
11\ 659\ 189.4\left( 5.4\right) \cdot10^{-10},\qquad[\tau].%
\end{array}
\right.   \label{AMMee}
\end{equation}
The difference between the experimental determination of $a_{\mu}$ and the
standard model using the $e^{+}e^{-}$ or $\tau$ data for the calculation of
the hadronic vacuum polarization (HVP) contribution is $3.6\sigma$ and $2.4\sigma$,
respectively.

The standard model prediction for $a_{\mu}$ consists of quantum
electrodynamics, weak and hadronic contributions. The QED and weak
contributions to $a_{\mu}$ have been calculated with great accuracy \cite{Aoyama:2012wk}
\begin{equation}
a_{\mu}^{\mathrm{QED}}=11~658~471.8951(0.0080)\cdot10^{-10}   \label{AMMqed}
\end{equation}
and \cite{Czarnecki:2002nt}
\begin{equation}
a_{\mu}^{\mathrm{EW}}=15.4(0.2)\cdot10^{-10}.   \label{AMMweak}
\end{equation}

The theoretical errors in (\ref{AMMee}) are
dominated by the uncertainties induced by the HVP and LbL effects.
Thus, to confront usefully theory with the experiment requires a better
determination of the hadronic contributions. In the last decade, a
substantial improvement in the accuracy of the contribution from the
HVP was reached. It uses, essentially, precise
determination of the low energy spectrum of the total $e^{+}e^{-}\rightarrow$
hadrons and inclusive $\tau$ lepton decays cross-sections. The HVP contributions at order $\alpha^{2}$ quoted in the most recent
articles on the subject are given in the Table.\\[0.1cm]

\vspace{-0.4cm}
Table.\\[0.1cm]
Phenomenological estimates and references for the leading order HVP contribution to the muon anomalous magnetic
moment based on $e^{+}e^{-}$ and $\tau$ data sets.\\[0.1cm]

\hspace{-0.5cm}
\begin{tabular}{|c|c|c|c|c|c|}
\hline
& $e^{+}e^{-}\cite{Davier:2010nc}$ & $\tau\cite{Davier:2010nc}$ & $e^{+}e^{-}\cite{Benayoun:2012wc}$ &
$e^{+}e^{-}\cite{Hagiwara:2011af}$  \\ \hline
$a_{\mu}^{\mathrm{HVP}~\left( 1\right) }\cdot10^{10}$ & $692.3\pm4.2$ & $%
701.5\pm4.7$ & $681.23\pm4.51$ & $694.91\pm4.27$  \\ \hline
\end{tabular}
\\[0.3cm]

The higher-order contributions at $O(\alpha^{3})$ level to $a_{\mu }^{%
\mathrm{HVP}~\left( 2\right) }$ was estimated in \cite{Hagiwara:2011af},
\begin{equation}
a_{\mu}^{\mathrm{HVP}~\left( 2\right) }=-9.84(0.07)\cdot10^{-10},
\label{AMM2}
\end{equation}
by using analytical kernel functions and experimental data on the $%
e^{+}e^{-}\rightarrow$ hadrons cross-section. In addition, there exists the $%
O(\alpha^{3})$ contribution to $a_{\mu}$ from the LbL diagram, $a_{\mu}^{\mathrm{hLbL}}$, that cannot be
expressed as a convolution of experimentally accessible observables and need
to be estimated from theory. In some works \cite{Prades:2009tw}, the value
\begin{equation}
a_{\mu}^{\mathrm{hLbL}}=10.5(2.6)\cdot10^{-10}   \label{AMM_LL}
\end{equation}
is considered as an estimate of the hadronic LbL contribution to the muon AMM.

\begin{figure}[h]
\begin{center}
\hspace{3.5cm}
\begin{tabular*}{\columnwidth}{@{}ccc@{}}
\raisebox{-0.5\height}{\resizebox{0.13\textwidth}{!}{\includegraphics{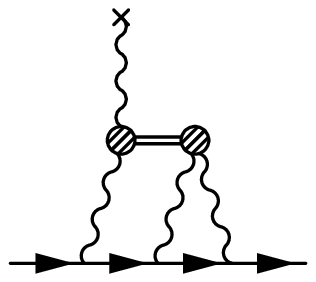}}}&
\raisebox{-0.5\height}{\resizebox{0.13\textwidth}{!}{\includegraphics{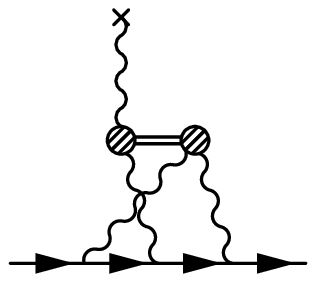}}}&
\raisebox{-0.5\height}{\resizebox{0.13\textwidth}{!}{\includegraphics{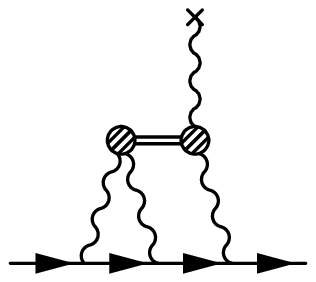}}}\\
(a)&(b)&(c)
\end{tabular*}
\end{center}
\caption{\emph{Hadronic LbL scattering contribution due to quark-antiquark exchanges.}}
\label{fig:LbL}
 \vspace{-0.5cm}
\end{figure}

Assuming absence of New Physics effects, a phenomenological estimate of the total hadronic contributions to $a_{\mu
}^{\mathrm{HVP} }$ has to be compared with the value deduced from the $g-2$
experiment (\ref{AMMg-2}) and known electroweak (\ref{AMMweak}) and QED (\ref{AMMqed}) corrections
\begin{equation}
a_{\mu}^{\mathrm{BNL}}-a_{\mu}^{\mathrm{QED}}-a_{\mu}^{\mathrm{EW}}=721.6\left( 6.3\right) \cdot10^{-10}.
\label{aMMH1EXP}
\end{equation}

\begin{wrapfigure}{l}{0.4\textwidth}
 \vspace{-1cm}  
  \begin{center}
    \includegraphics[width=0.4\textwidth]{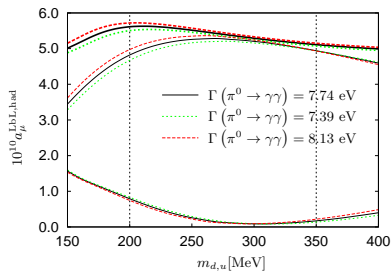}
  \end{center}
 \vspace{-0.5cm}  
  \caption{\emph{ LbL contribution to the muon AMM
from the neutral pion and $\protect\sigma $ exchanges as a function of the
dynamical quark mass. Bunch of three lower lines correspond to the $\protect%
\sigma $ contribution, the $\protect\pi ^{0}$ contribution is in the middle,
and the upper lines are the combined contribution.  }}
\vspace{-0.1cm}
\label{fig:PionLbL}
\end{wrapfigure}

Two new experiments on measurement of $a_{\mu}$ are proposed at Fermilab (E989)\cite{Venanzoni:2012yp} and  J-PARC\cite{Saito:2012zz}
which plan to improve the experimental uncertainty by a factor of 4-5 with respect to the previous BNL experiment.
In that respect theoretical predictions of HVP and LbL contributions to the muon AMM should be at the same level or better than a precision of planed experiments. In next part we discuss the hadronic LbL contribution as it is calculated within the nonlocal chiral quark model (N$\chi$QM) of low energy QCD and show that, within this framework, it might be possible realistically to determine this value to a sufficiently safe accuracy. We want to discuss how well this model (see, e.g., \cite{Anikin:2000rq,Dorokhov:2003kf}) does in calculating $a_{\mu}^{\mathrm{LbL}}$.

\section{LbL contribution to the muon AMM due to light quark-antiquark exchanges in pseudoscalar and scalar channels}

The uncertainties of the SM value for $a_{\mu}$ are dominated by the uncertainties of the hadronic contributions, $a_{\mu}^{\mathrm{Strong}%
},$ since their evaluation involve quantum chromodynamics (QCD) at long-distances for which perturbation theory cannot be employed.
Below we discuss with some details theoretical status of hadronic LbL contribution to the muon AMM due to exchange by light mesons within N$\chi$QM.

N$\chi$QM is an effective model that has a numerous applications for description of low energy hadronic dynamics. We mention only those applications that are related to the problem of hadronic contributions to the muon AMM. The two-point VV correlator has been calculated in \cite{Dorokhov:2003kf} and used for calculations of the HVP contribution to the muon AMM \cite{Dorokhov:2004ze}. The three-point VAV correlator has been calculated in \cite{Dorokhov:2005pg,Dorokhov:2005hw} and used for calculations of the photon-$Z$-boson interference contribution to the muon AMM \cite{Dorokhov:2005ff}.

More recently the LbL contribution due to exchange of pseudoscalar (P) and scalar (S) mesons (Fig. \ref{fig:LbL}) was elaborated in \cite{Dorokhov:2008pw,Dorokhov:2011zf,Dorokhov:2012qa}. The vertices containing the virtual meson $M$ with momentum $p$ and two photons with momenta $q_{1,2}$ and the polarization vectors $\epsilon_{1,2}$
can be written as \cite{Bartos:2001pg}%
\begin{equation}
\mathcal{A}\left(  \gamma^{\ast}_{\left(  q_{1},\epsilon_{1}\right)}  \gamma^{\ast}_{\left( q_{2},\epsilon_{2}\right)}  \rightarrow M^{\ast}_{\left(  p\right)} \right)=e^{2}\epsilon_{1}^{\mu}\epsilon_{2}^{\nu}\Delta^{\mu\nu}_M\left( p, q_{1}, q_{2}\right)
\end{equation}
with
\begin{equation}
\Delta_P^{\mu\nu}\left( p, q_{1}, q_{2}\right)=-i\varepsilon_{\mu\nu\rho\sigma}q_{1}^{\rho}q_{2}^{\sigma} \mathrm{F}_{P}\left(  p^{2};q_{1}^{2},q_{2}^{2}\right) ,
\end{equation}
and
\begin{equation}
\Delta_S^{\mu\nu}\left( p, q_{1}, q_{2}\right)=
\mathrm{A}_{S}\left(p^{2};q_{1}^{2},q_{2}^{2}\right) P_{A}^{\mu \nu }(q_{1},q_{2})+
\mathrm{B'}_{S}(p^{2};q_{1}^{2},q_{2}^{2})P_{B'}^{\mu \nu }(q_{1},q_{2})
,\label{ApiGG}
\end{equation}
where
\begin{eqnarray}
&&\quad P_{A}^{\mu \nu }(q_{1},q_{2}) =\left( g^{\mu \nu }(q_{1}
q_{2})-q_{1}^{\nu }q_{2}^{\mu }\right) ,\nonumber \\
&&\quad P_{B'}^{\mu \nu }(q_{1},q_{2}) =\frac{\left( q_{1}^{2}q_{2}^{\mu }-(q_{1}
q_{2})q_{1}^{\mu }\right) \left( q_{2}^{2}q_{1}^{\nu }-(q_{1}
q_{2})q_{2}^{\nu }\right)}{(q_{1} q_{2})^{2}-q_{1}^{2}q_{2}^{2}} ,\nonumber
\end{eqnarray}
and  $p=q_{1}+q_{2}$. The subject of model calculations\footnote{For details see \cite{Dorokhov:2012qa}} are the (P/S)VV vertice functions $\mathrm{F}_{P},\mathrm{A}_{S},\mathrm{B'}_{S}$.

The expression for the LbL contribution
to the muon AMM from the light meson exchanges can be written as
\begin{eqnarray}
&& a_{\mu}^{\mathrm{LbL},\mathrm{M}}=-\frac{4\alpha^{3}}{3\pi^{2}}%
\int\limits_{0}^{\infty}dQ_{1}^{2}\int\limits_{0}^{\infty}dQ_{2}^{2}\int%
\limits_{-1}^{1}dt\sqrt{1-t^{2}}\frac{1}{Q_{3}^{2}} \sum_{M} \biggl[ \frac{\mathcal{N}^{M}_1(Q_1^2,Q_2^2,Q_3^2)}{Q_{2}^{2}+\mathrm{M}_{M}^{2}} +\frac{\mathcal{N}^{M}_2(Q_1^2,Q_3^2,Q_2^2)}{2(Q_{3}^{2}+\mathrm{M}_{M}^{2})} \biggr] ,   \nonumber\\
&&\quad \mathcal{N}^{P}_{\mathbf{1,2}}(Q_1^2,Q_2^2,Q_3^2)=
  \mathrm{F}_{P}\left( Q_{2}^{2} ;Q_{2}^{2},0\right)\mathrm{F}_{P}\left( Q_{2}^{2};Q_{1}^{2},Q_{3}^{2}\right)\mathrm{Tp}_{\mathbf{1,2}},\nonumber\\
&&\quad \mathcal{N}^{S}_{1,2}(Q_1^2,Q_2^2,Q_3^2)=\left(A\left( Q_{2}^{2} ;Q_{2}^{2},0\right)+\frac{1}{2}B'\left( Q_{2}^{2} ;Q_{2}^{2},0\right)\right)\nonumber\\
&&\times\left(A\left( Q_{2}^{2};Q_{1}^{2},Q_{3}^{2}\right)\mathrm{Ts}^{\mathrm{AA}}_{\mathbf{1,2}}+
\frac{1}{2}B\left( Q_{2}^{2};Q_{1}^{2},Q_{3}^{2}\right)\mathrm{Ts}^{\mathrm{AB}}_{\mathbf{1,2}}\right), \nonumber
\end{eqnarray}
where $Q_{3}=-\left( Q_{1}+Q_{2}\right)$ and $B_S=B'_S/((q_{1} q_{2})^{2}-q_{1}^{2}q_{2}^{2})$. The kinematic factors $\mathrm{Tp}_{\mathbf{i}}$ and $\mathrm{Ts}_{\mathbf{i}}$ can be found in \cite{Knecht:2001qf} and \cite{Dorokhov:2012qa}, respectively.

\section{The results of model calculations}

\begin{wrapfigure}{l}{0.49\textwidth}
 \vspace{-0.5cm}  
  \begin{center}
    \includegraphics[width=0.48\textwidth]{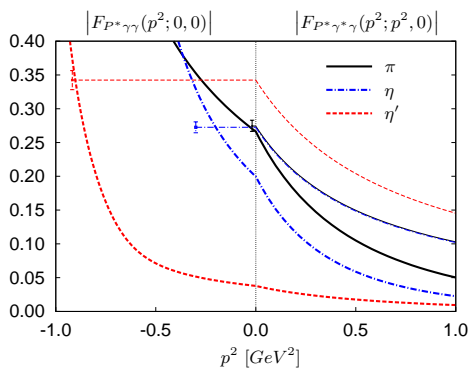}
  \end{center}
   \vspace{-0.5cm}  
  \caption{ \emph{Plots of the $\pi^{0},\eta$ and $\eta^{\prime}$ vertices $F_{P^{\ast}\gamma\gamma}(p^{2};0,0)$ in the timelike
region and $F_{P^{\ast}\gamma^{\ast}\gamma}(p^{2};p^{2},0)$ in the spacelike
region in N$\chi$QM model (thick lines) and VMD model (thin
lines). The points with error bars correspond to the physical points of the
meson decays into two photons. The VMD curves for $\pi^{0}$ and $\eta$ are
almost indistinguishable.}}
\vspace{0.15cm}  
\label{fig:CompFF}
\end{wrapfigure}

The total contribution of pseudoscalar ($\pi^0,\eta,\eta'$) exchanges is estimated as
\begin{equation}
a_{\mu }^{\mathrm{LbL,PS}}=(5.85\pm 0.87)\cdot 10^{-10},
\end{equation}
and the combined value for the
scalar ($\sigma,a_0,f_0$) and pseudoscalar contributions is \cite{Dorokhov:2012qa} 
\begin{equation}
a_{\mu }^{\mathrm{LbL,PS+S}}=(6.25\pm 0.83)\cdot 10^{-10}.
\end{equation}

We found that within the N$\chi $QM the pseudoscalar meson contributions to the
muon AMM are systematically lower then the results
obtained in the other works. The
full kinematic dependence\footnote{This dependence also recently studied in \cite{Goecke:2010if}.} of the vertices on the pion virtuality diminishes
the result by about 20-30\% as compared to the case where this dependence is
neglected. For $\eta $ and $\eta ^{\prime }$ mesons the results are reduced
by factor about 3 in comparison with the results obtained in other models
where the kinematic dependence was neglected (see Fig. \ref{fig:CompFF} and discussion in \cite{Dorokhov:2011zf,Dorokhov:2012qa}). The scalar mesons
contribution is small positive and partially compensates model dependence of the
pseudoscalar contribution (Fig. \ref{fig:PionLbL}). 

\section{Acknowledgements}

A.E.D. thanks Wojtek Broniowski for nicely organized Light Cone meeting. This work is supported in part by the Bogoliubov-Infeld program (JINR),
the Russian Foundation for Basic Research (projects Nos.~10-02-00368, 11-02-00112 and 12-02-31874),
the Federal Target Program Research and Training Specialists
in Innovative Russia 2009-2013 (16. 740.11.0154).

\end{document}